\documentclass[prb,preprint,showpacs,preprintnumbers,amsmath,amssymb]{revtex4-1}

\usepackage[utf8]{inputenc}
\usepackage{amsmath, amssymb,graphicx}
\usepackage[cdot,mediumqspace,squaren]{SIunits}
\usepackage{color}
\usepackage{epsfig}

\renewcommand{\vec}{\mathbf} % Vecteur
\newcommand{\e}[1]{\text{e}^{#1}} % exponentielle
\renewcommand{\Im}[1]{\text{Im}(#1)} % partie reelle
\newcommand{\eps}{ \varepsilon} % epsilon
\newcommand{\Einc}{E_\text{inc}} % incident field

\begin{document}

\title{Optical meta-atom for localization of light with quantized energy}

\author{Sylvain Lanneb\`{e}re}
\author{M\'{a}rio G. Silveirinha}
\email{To whom correspondence should be addressed:
mario.silveirinha@co.it.pt}
\affiliation{University of Coimbra,
Department of Electrical Engineering -- Instituto de
Telecomunica\c{c}\~{o}es, Coimbra 3030-290, Portugal}

\date{\today}

\begin{abstract}
The capacity to confine light into a small region of space is of paramount importance in
many areas of modern science.
Here, we suggest a mechanism to store a quantized ``bit'' of light
-- with a very precise amount of energy -- in an open core-shell
plasmonic structure (``meta-atom'') with a nonlinear optical
response. Notwithstanding the trapped light state is embedded in the
radiation continuum, its lifetime is not limited by the radiation
loss. Interestingly, it is shown that the interplay between the
nonlinear response and volume plasmons enables breaking fundamental
reciprocity restrictions, and coupling very efficiently an external
light source to the meta-atom. The collision of an incident optical
pulse with the meta-atom may be used to release the trapped
radiation ``bit''.

\end{abstract}

%\pacs{42.70.Qs, 78.67.Pt, 03.65.Ge, 73.20.Mf}

\date{\today}

\maketitle
%%%%%%%%%%%%%%%%%%%%%%%%%%%%%%%%%%%%%%%%%%%%%%%%%%%%%%%%%%%%%%%%%%%%%%%%%%%%%%%%%%%%%%%%%%%%%%%%%%%%%%%%%%%%%%%%%%%%%%%%%%%%%%%%%%%%%%%%%%%%%%%%%%%%%%%%%%%%%%%
%%%%%%%%%%%%%%%%%%%%%%%%%%%%%%%%%%%%%%%%%%%%%%%%%%%%%%%%%%%%%%%%%%%%%%%%%%%%%%%%%%%%%%%%%%%%%%%%%%%%%%%%%%%%%%%%%%%%%%%%%%%%%%%%%%%%%%%%%%%%%%%%%%%%%%%%%%%%%%%

Light localization has several important technological applications
\cite{vahala_optical_2003}, and is usually achieved with the help of
physical barriers such as mirrors \cite{rempe_measurement_1992} and
photonic band-gap materials
\cite{lalanne_photon_2008,sekoguchi_photonic_2014} that act to
strongly reduce the effects of radiation loss. Light is however an
object difficult to tame: no matter how elaborate and intricate are
the material constructions that may be used to screen it from the
exterior environment there is always some residual coupling with the
radiation continuum, and hence light - if not absorbed by the
material walls - always finds its way out. Indeed, in any
conventional open resonator (e.g. whispering gallery resonators
\cite{matsko_optical_2006,grudinin_ultrahigh_2006}, or metallic
nanoparticles \cite{Tribelsky,Lukyanchuk}) the coupling to the
surrounding region is never totally suppressed, and radiation loss
is one of the factors that limits the lifetime of light
oscillations. Even though the radiation emission due to specific
electric current oscillation modes may be residual (e.g. the anapole current
distribution \cite{miroshnichenko_seeing_2014}), it
is never precisely zero.

In the last decades, there has been a great interest in alternative
mechanisms to localize light within the radiation continuum
\cite{lagendijk_fifty_2009,
marinica_bound_2008,plotnik_experimental_2011,lee_observation_2012,hsu_observation_2013,monticone_embedded_2014,lepetit_controlling_2014,zhen_topological_2014}.
An interesting approach is based on an old idea by von Neumann and
Wigner who discovered that certain  electric potentials may support
spatially localized electron states with energies higher than the
potential barriers \cite{Neumann_1929,stillinger_bound_1975}. In
recent years, different groups have extended this idea to light
waves, and it has been shown that open material structures with
tailored geometries may support localized light states \cite{
marinica_bound_2008,plotnik_experimental_2011,lee_observation_2012,hsu_observation_2013,lepetit_controlling_2014,zhen_topological_2014}.
In these structures, the discrete light spectrum overlaps the
continuous spectrum, and hence these spatially localized states with
infinite lifetimes are technically known as ``embedded
eigenvalues''. The existence of such states is highly nontrivial and
truly remarkable, because it shows that light may be localized based
simply on the scattering provided by a set of surrounding
transparent (ideally lossless) material objects. There is however a
downside: until recently, the known solutions were based on
infinitely extended material profiles, e.g. a photonic crystal.
Thus, the objects that localize the radiation are required to be
placed also at arbitrarily large distances from the spot wherein the
light is concentrated. If the structure is truncated the
localization becomes imperfect, and the oscillation lifetime becomes
finite.

Recently, we introduced a novel approach to trap light in a
bounded open cavity with suppressed radiation loss
\cite{silveirinha_trapping_2014}. It was theoretically shown that
under some strict conditions, volume plasmons -- i.e. charge density
waves in metals -- may enable the formation of ``embedded
eigenvalue'' states in finite sized cavities, such that in the limit
of no material loss the light oscillations can have infinitely long
lifetimes. Our proposal applies to a wide range of resonators, and
in particular the light volume may be ultra-subwavelength
\cite{monticone_embedded_2014, silveirinha_trapping_2014}. In a
two-layer spherical structure, the shell is ideally formed by a
material with vanishing permittivity $\varepsilon=0$ -- so that it
supports volume plasmons
 -- and the core region is a vacuum or a standard
dielectric \cite{silveirinha_trapping_2014}.

A limitation of this system is that because of the Lorentz
reciprocity theorem the trapped light state cannot be pumped by an
external source. Indeed, structures made of reciprocal materials are
intrinsically bi-directional, and hence if the trapped light cannot
leak out then it is also impossible to feed its oscillations with an
external excitation \cite{silveirinha_trapping_2014}. Here, it is
shown that the interplay of volume plasmons with a nonlinear
material response may provide the means to pump the embedded
eigenstate using an external source. Notably, it is proven that the
energy associated with the embedded eigenvalue is quantized, such
that self-sustained oscillations are only possible for specific
stored energy values. Because of the obvious parallelisms with the
energy quantization of bound electronic states in atoms, we refer to
the proposed resonator as an ``optical meta-atom''. In realistic
systems, the material absorption limits the number of cycles during
which the radiation can be trapped in the optical meta-atom. In
principle, this limitation may be compensated using a gain medium.
It is envisioned that a gain compensated meta-atom may be used as an
elementary one-bit optical memory.

\section*{Results}

\subsection*{The meta-atom}
\begin{figure}[!tbh]
\centering \epsfig{file=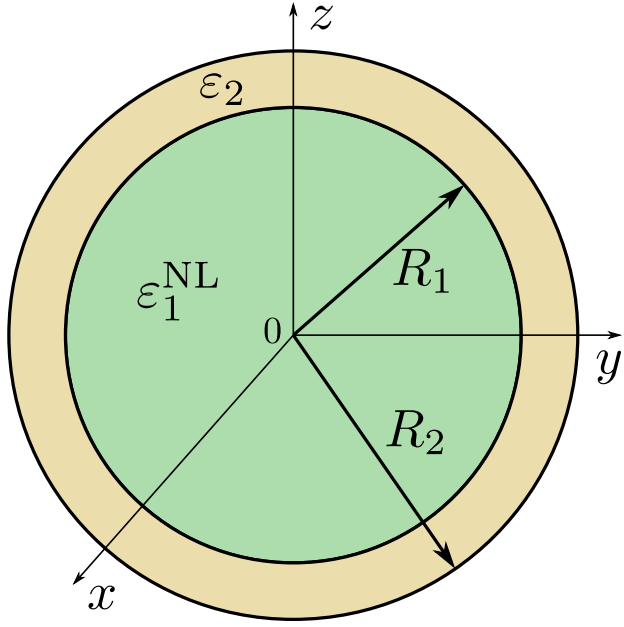, width=0.5\linewidth}
         \caption{\textbf{The optical meta-atom.} The core material is a dielectric with a
         nonlinear permittivity response $\eps_1^{\rm{NL}}$, while the shell is a plasmonic material with permittivity $\eps_2(\omega)$.
         The inner  and outer radii are $R_1$ and $R_2$, respectively.
          }
\label{fig:dessin_core-shell_particle}
\end{figure}
The geometry of the core-shell spherical optical meta-atom is
represented in Fig. \ref{fig:dessin_core-shell_particle}. The shell
 has radius $R_2$, and, without loss of generality, its relative permittivity is supposed to follow
a Drude dispersion model $\eps_2(\omega)=
1-\frac{\omega_\text{p}^2}{\omega(\omega+i\omega_\text{c})}$, where $\omega_\text{p}$ is
the plasma frequency and $\omega_\text{c}$ is the collision frequency. The
core has radius $R_1$ and is made of a dielectric material with
relative permittivity $\eps_1$. As shown in our previous work, the
volume plasmons in the shell at $\omega = \omega_\text{p}$ may perfectly
screen the light in the core region, so that in the limit of
vanishing material loss ($\omega_\text{c} \to$ 0) the oscillations lifetime
may be infinite \cite{silveirinha_trapping_2014}. This effect
requires that at $\omega=\omega_\text{p}$ one has $j_n(k_1R_1)=0$, where
$j_n$ is the spherical Bessel function, $k_1= {\omega_\text{p} \sqrt
{\varepsilon _1 } /c}$, and $n \ge 1$ is the azimuthal quantum
number. For a fixed $\eps_1$, this condition is satisfied only for
certain specific values of the core radius. For instance, for $n=1$
the trapped fields have a dipolar ($p$-type) symmetry and the first
zero of $j_1(u)$ occurs at $u\approx4.49$, so that the corresponding
optimal radius is $R_{1,0}=4.49 c/(\omega_\text{p}\sqrt{\eps_1})$
\cite{silveirinha_trapping_2014}. Interestingly, in the ideal regime
$\eps_2=0$ the radiation is held within the core forever, even for a deeply subwavelength shell. This contrasts with
conventional high-Q optical cavities. For instance, we numerically
verified (not shown) that to achieve a quality factor of the order
of 100 in a Fabry-Perot cavity with lossless photonic crystal walls,
the mirror walls need to be at least six wavelengths thick.
Importantly, any perturbation of the optimal radius, no matter how
small, will lead to a finite oscillation lifetime
\cite{silveirinha_trapping_2014}. Moreover, as discussed previously,
even if the inner radius could be tuned to exactly satisfy
$R_1=R_{1,0}$, it would still be impossible to externally pump the
light oscillations.

To overcome these restrictions, here we investigate the
opportunities created by a core region with an optical nonlinearity.
It is assumed that the core material has an instantaneous isotropic
Kerr response described by the nonlinear third order susceptibility
$\chi^{(3)}$ \cite{boyd_nonlinear_2008}, such that the electric
displacement vector satisfies ${\bf{D}}(t) = \varepsilon _0
\varepsilon ^{{\rm{NL}}} {\bf{E}}(t)$, with $\varepsilon
^{{\rm{NL}}} = \varepsilon_1  + \chi ^{\left( 3 \right)} \left|
\vec{E}(t) \right|^2$  where $\varepsilon_1$ is the relative dielectric
permittivity for weak field intensities. The key idea is to choose
an inner radius $R_1$ slightly different from $R_\text{1,0}$ (so
that the field oscillations can be externally pumped), and take
advantage of the nonlinear dynamics to self-tune the resonator.
Heuristically, one may expect that when
$k_1^\text{NL}R_1=k_1R_{1,0}$ the radiation loss may be strongly
suppressed, where we put $k_1^\text{NL} = \sqrt {\varepsilon
^{{\rm{NL}}} } \omega_\text{p} /c$. This condition is equivalent to:
\begin{equation}
\chi ^{\left( 3 \right)} \frac{3}{4}\left| {{\bf{E}}_{\omega_\text{p} }}
\right|^2 =\eps_{1}\left[ \left(\frac{R_{1,0}}{ R_1} \right)^2 - 1
\right].
 \label{E:expected_value_Chi_times_Esquare}
\end{equation}
Note that if ${\bf{E}}\left( t \right) \approx {\mathop{\rm
Re}\nolimits} \left\{ {{\bf{E}}_\omega e^{ - i\omega t} } \right\}$
then
 ${\bf{D}}\left( t
\right) \approx {\mathop{\rm Re}\nolimits} \left\{ {\varepsilon _0
\left( {\varepsilon _1  + \frac{3}{4}\chi ^{\left( 3 \right)} \left|
{{\bf{E}}_\omega  } \right|^2 } \right){\bf{E}}_\omega  e^{ -
i\omega t} } \right\}$ for a weak nonlinearity. This result was used
to estimate $\varepsilon ^{{\rm{NL}}}$ in the core region. Hence,
for a self-focusing Kerr material ($\chi ^{\left( 3 \right)}>0$) the
resonator self-tuning may be feasible when $R_1<R_{1,0}$.
\subsection*{Temporal dynamics of the electromagnetic field}
To put these ideas on a firm ground, next we develop a simple
analytical model for the nonlinear dynamics of the core electric
field in a scenario wherein the optical meta-atom is illuminated by
an incoming plane wave. To begin with, we consider the linear case
($\chi ^{\left( 3 \right)}=0$) and note that under plane wave
incidence, the electric field at the center of the core-shell
particle is given in the frequency domain by $\left. \vec{E}
\right|_{\vec{r}=\vec{0}}= a_1^\text{TM} E_{\rm{inc}} \hat{\vec{x}}$
\cite{silveirinha_trapping_2014,bohren}. Here,  ${\bf{E}}_{\rm{inc}}
= \Einc \left( \omega \right){\bf{\hat x}}$ is the linearly
polarized incident field calculated at the center of the particle
and $a_1^\text{TM}$ is the first order Mie coefficient for
transverse radial magnetic (${\rm{TM}}^r$) waves. It was shown in
\cite{silveirinha_trapping_2014} that for $R_1 \approx R_{1,0}$ and
$\omega \approx \omega_\text{p}$ the Mie coefficient $a_1^\text{TM}$
satisfies
\begin{equation}
 a_1^\text{TM}\approx \e{i\phi_0} \frac{(\omega-\omega_\text{L})(\omega+\omega_\text{L}^*)}{(\omega-\omega_\text{r})(\omega+\omega_\text{r}^*) }
 \approx
 e^{i\phi _0 } \frac{{\omega  - \omega_\text{L} }}{{\omega  - \omega_\text{r} }}
 ,
 \label{E:approx_a1}
\end{equation}
where $\phi_0(\omega)$ is some phase factor associated with a time
delay, and the second identity assumes that both $\omega_\text{L}$ and
$\omega_\text{r}$ are near $\omega_\text{p}$. In the above, $\omega_\text{L} = \omega'_\text{L}
+ i \omega''_\text{L} $ is the complex frequency associated with plasmon
oscillations, which is defined by $\eps_2(\omega_\text{L})=0$. Moreover,
$\omega_\text{r}  = \omega'_\text{r} + i\omega''_\text{r} $ is the complex resonance
frequency of the trapped mode, and it may be numerically calculated
as explained in \cite{silveirinha_trapping_2014}. When $\omega =
\omega'_\text{r}$ the inner field may be strongly enhanced due to the
excitation of a trapped state with quality factor $Q = \omega'_\text{r}
/\left( { - 2\omega''_\text{r} } \right)$, whereas when $\omega = \omega
'_\text{L}$ it is near zero due to the screening provided by the volume
plasmons. The response has a Fano-type lineshape \cite{Lukyanchuk}.

If the spectrum of the incident field is concentrated at $\omega_\text{p}$,
it is possible to write in the time-domain $ E_{{\rm{inc}}} \left( t
\right) = {\mathop{\rm Re}\nolimits} \left\{ {E_{\omega_\text{p}
}^{{\rm{inc}}} \left( t \right)e^{ - i\omega_\text{p} t} } \right\}$ and
$E\left( t \right) = {\mathop{\rm Re}\nolimits} \left\{ {E_{\omega_\text{p} }^{{\rm{tot}}} \left( t \right)e^{ - i\omega_\text{p} t} } \right\}$,
with the envelopes of the incident field ${E_{\omega_\text{p}
}^{{\rm{inc}}} \left( t \right)}$ and of the total field ${E_{\omega_\text{p} }^{{\rm{tot}}} \left( t \right)}$ varying slowly in time.
Calculating the inverse Fourier transform of $E = a_1^\text{TM}
\Einc$, it is readily found that the differential equation governing
the time evolution of the total field envelope at the center of the
core-shell particle is:
\begin{eqnarray}
\left( {i\frac{\partial }{{\partial t}} + \omega_\text{p}  - \omega_\text{r} }
\right)E_{\omega_\text{p} }^{{\rm{tot}}} \left( t \right) = \left(
{i\frac{\partial }{{\partial t}} + \omega_\text{p}  - \omega_\text{L} }
\right)E_{\omega_\text{p} }^{{\rm{inc}}} \left( t \right). \nonumber \\
 \label{E:differential_equation_time}
\end{eqnarray}
For simplicity the phase factor $\phi_0$ was dropped.

To validate this theory, next it is assumed that the inner core has
a linear response with permittivity $\eps_1=1$ and that the
plasmonic shell has radius $R_2=1.1 R_1$ described by a Drude model
with the plasma frequency $\omega_\text{p}/2\pi= 750 ~\tera\hertz$
(violet light) and negligible material loss. For this configuration
the optimal radius is $R_{1,0} \approx 286 ~\nano\meter$. The
core-shell particle is excited by an $x$-polarized incident field
with a Gaussian profile in the time domain with envelope $
E^\text{inc}_{\omega_\text{p}}(z,t)= E_0 e^{i \frac{\omega_\text{p}}{c} z}
e^{-\left(\frac{t-t_0}{\sigma}\right)^2}$, propagating along the
$z$-direction. The time duration of the pulse is determined by the
full-width-half-maximum (\textit{fwhm}) $\Delta
\tau=2\sigma\sqrt{\ln{2}}$, and $t_0$ is the time instant for which
the incoming wave field is peaked.
\begin{figure}[!h]
\centering \epsfig{file=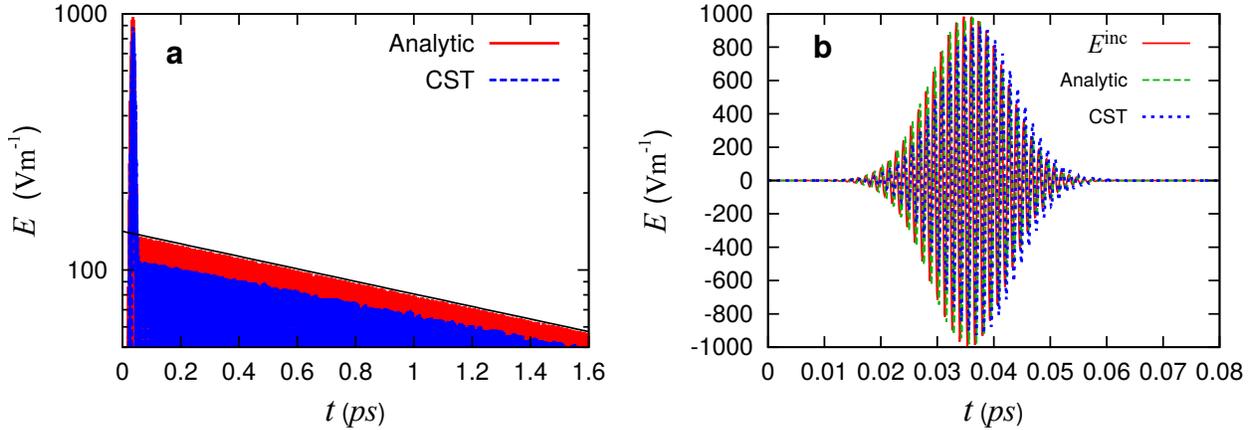, width=1\linewidth}
         \caption{\textbf{Excitation of the meta-atom in the linear regime.} Electric field
         at the center of the meta-atom as a function of time. The
         incident pulse duration is $\Delta \tau=16.3 ~\femto\second$.
         (\textbf{a}) $R_1 \approx 0.98 R_{1,0}$.
         The black thin line represents the theoretical peak amplitude determined by the decay rate $\omega''_\text{r}$.
         (\textbf{b})  $R_1=R_{1,0}$.
} \label{fig:linear_results}
\end{figure}
First, we consider that $R_1 \approx 0.98 R_{1,0}$, and suppose that
the incident pulse has a peak amplitude $E_0=1000~\volt\meter^{-1}$,
and a duration $\Delta \tau=16.3 ~\femto\second$. Figure
\ref{fig:linear_results}a represents in a semi-logarithmic scale the
$x$-component of the electric field at the center of the meta-atom
as a function of time. The solid red line was calculated using the
analytical model \eqref{E:differential_equation_time} and the dashed
blue line was obtained with CST Microwave Studio \texttrademark
\cite{CST}. As seen, apart from a small amplitude shift, the
analytic and full wave results are rather similar. This confirms
that the differential equation \eqref{E:differential_equation_time}
may be used to characterize the dynamics of the electric field in
the core region. Remarkably, even though the sphere's diameter
comparable with the wavelength, the temporal evolution of the field
inside the meta-atom is well described by the electric dipolar mode.
Indeed, the meta-atom is designed to exclusively trap this
oscillation mode and therefore the other modes, even if present,
decay quickly after the initial excitation period. In the example
under study, due to the relatively small size of the meta-atom the
secondary resonance nearer $\omega_\text{p}$ occurs at $\omega \approx 0.8
\omega_\text{p}$ and has a quality factor ten times smaller than that
associated with the primary resonance. This low quality factor is
justified by the fact that the shell becomes weakly reflecting when
the oscillation frequency drops to $0.8 \omega_\text{p}$ .

Notably, after the incident pulse overtakes the meta-atom, the field
continues to oscillate inside the core with a peak oscillation
amplitude that decays exponentially. The peak amplitude is well
described by the formula $A_0 \e{\omega''_\text{r} t}$ (black line in Fig.
\ref{fig:linear_results}a), where $A_0$ is some fitting constant and
$\omega''_\text{r}=\Im{\omega_\text{r}} \simeq -1.2\cdot10^{-4} \omega_\text{p}$ is the
decay rate calculated with the theory of Ref.
\cite{silveirinha_trapping_2014}. Note that the free oscillation
amplitude is significantly less than $E_0$ and, very importantly, it
is strictly linked to the free oscillation decay rate. To show this,
we plot in Fig. \ref{fig:linear_results}b the electric field at the
center of a meta-atom with $R_1 = R_{1,0}$. The incident field is
the same as in the previous example. Crucially, the field inside the
meta-atom vanishes almost instantaneously after the end of the
incident pulse, notwithstanding that for this configuration the
theoretical decay rate vanishes: $\omega''_\text{r} = 0$. This result
confirms that for a perfectly tuned cavity it is impossible to pump
the ``embedded eigenvalue'' state with an external excitation, in
agreement with the Lorentz reciprocity theorem
\cite{silveirinha_trapping_2014}. Indeed, in the linear regime,
increasing the trapped state lifetime (i.e. decreasing the scattered
power) inevitably implies decreasing the external coupling strength
(i.e the extracted power see
\cite{tretyakov_maximizing_2014,liberal_least_2014}).

\subsection*{Trapping a light bit}

To show how one may take advantage of the nonlinear dynamics to
surpass these fundamental limitations, next Eq.
\eqref{E:differential_equation_time} is generalized to include the
effect of a nonlinear material response. To do this, $\omega_\text{r}$ is
regarded as a function of the inner core permittivity $\omega_\text{r} =
\omega_\text{r} \left( {\varepsilon ^{\rm{NL}} } \right) \approx \omega_\text{p}
+ i\,\omega''_\text{r} \left( {\varepsilon ^{\rm{NL}} } \right)$. The
decay rate $-\omega''_\text{r}$ has a minimum ($- \omega''_\text{r} = \omega
''_{\min }$) for the optimal inner core permittivity $\varepsilon
_{1,\text{opt}} \approx \varepsilon _1 \left( {R_{1,0} /R} \right)^2$. The
minimum is proportional to the decay rate of the volume plasmon
oscillations: $\omega ''_{\min } \approx \omega_\text{c} /2$
\cite{silveirinha_trapping_2014}. Thus, one may write:
\begin{eqnarray}
\omega_\text{r}  \approx \omega_\text{p}  - i\,\left[ {\omega ''_{\min }  +
\alpha \left( {\varepsilon ^{{\rm{NL}}}  - \varepsilon _{1,\text{opt}} }
\right)^2 } \right], \label{E:omega_r_approx}
\end{eqnarray}
where $\varepsilon ^{{\rm{NL}}}  = \varepsilon _1  + \frac{3}{4}\chi
^{\left( 3 \right)} \left| {E_{\omega_\text{p} }^{{\rm{tot}}} }
\right|^2$, and $\alpha$ is some constant that can be obtained from
a Taylor expansion of the formula $\omega_\text{r}=\omega_\text{r}(\varepsilon_1)$
around $\varepsilon_1  = \varepsilon _{1,\text{opt}} $. For example, for
$R_2=1.1 R_1$, and $R_1 = 0.98 R_{1,0}$
%$R_1 \omega_\text{p}/c = 0.98 \times 4.49$
it is found that $\alpha \approx 0.072 \omega_\text{p}$. Substituting Eq.
\eqref{E:omega_r_approx} into Eq.
\eqref{E:differential_equation_time} it is possible to characterize
the nonlinear dynamics of the system. Importantly, for time
intervals wherein the incident field vanishes (e.g. after the
incoming pulse overtakes the meta-atom) the inner field envelope is
determined by $\frac{{\partial E_{\omega_\text{p} }^{\text{tot}} }}{{\partial t}}
=  - \left[ {\omega ''_{\min } + \alpha \left( {\varepsilon
^{{\rm{NL}}}  - \varepsilon _{1,\text{opt}} } \right)^2 } \right]E_{\omega_\text{p} }^{\text{tot}} $. Hence, if the plasmons decay rate is negligible or if
it is compensated by some gain mechanism ($\omega ''_{\rm{min}}=0$),
it is possible to suppress the radiation loss when $\varepsilon
^{{\rm{NL}}}  = \varepsilon _{1,\text{opt}}  = \varepsilon _1 \left(
{R_{1,0} /R} \right)^2$. This condition is equivalent to Eq.
\eqref{E:expected_value_Chi_times_Esquare}, and implies that the
stored field peak amplitude is precisely determined by the strength
of the nonlinearity. Thus, a nonlinear material response may enable
us to efficiently pump the embedded ``eigenstate'' with an external
source, such that the trapped radiation can be retained within the
resonator for an extremely long time, only limited by the material
loss in the meta-atom. Moreover, Eq.
\eqref{E:expected_value_Chi_times_Esquare} suggests that the amount
of energy that may be retained within the optical meta-atom should
be independent of the excitation. Thus, the energy of a trapped
radiation ``bit' is precisely quantized, and in principle can only
have a single nonzero value determined by the value of $\chi^{(3)}$.
For $n=1$, the trapped state is associated with dipolar-type
oscillations and hence it is triply degenerate in the linear regime
\cite{silveirinha_trapping_2014}. Thus, the optical meta-atom may be
a photonic analogue of an atomic system with degenerate
energy levels.

\begin{figure}[!th]
\centering \epsfig{file=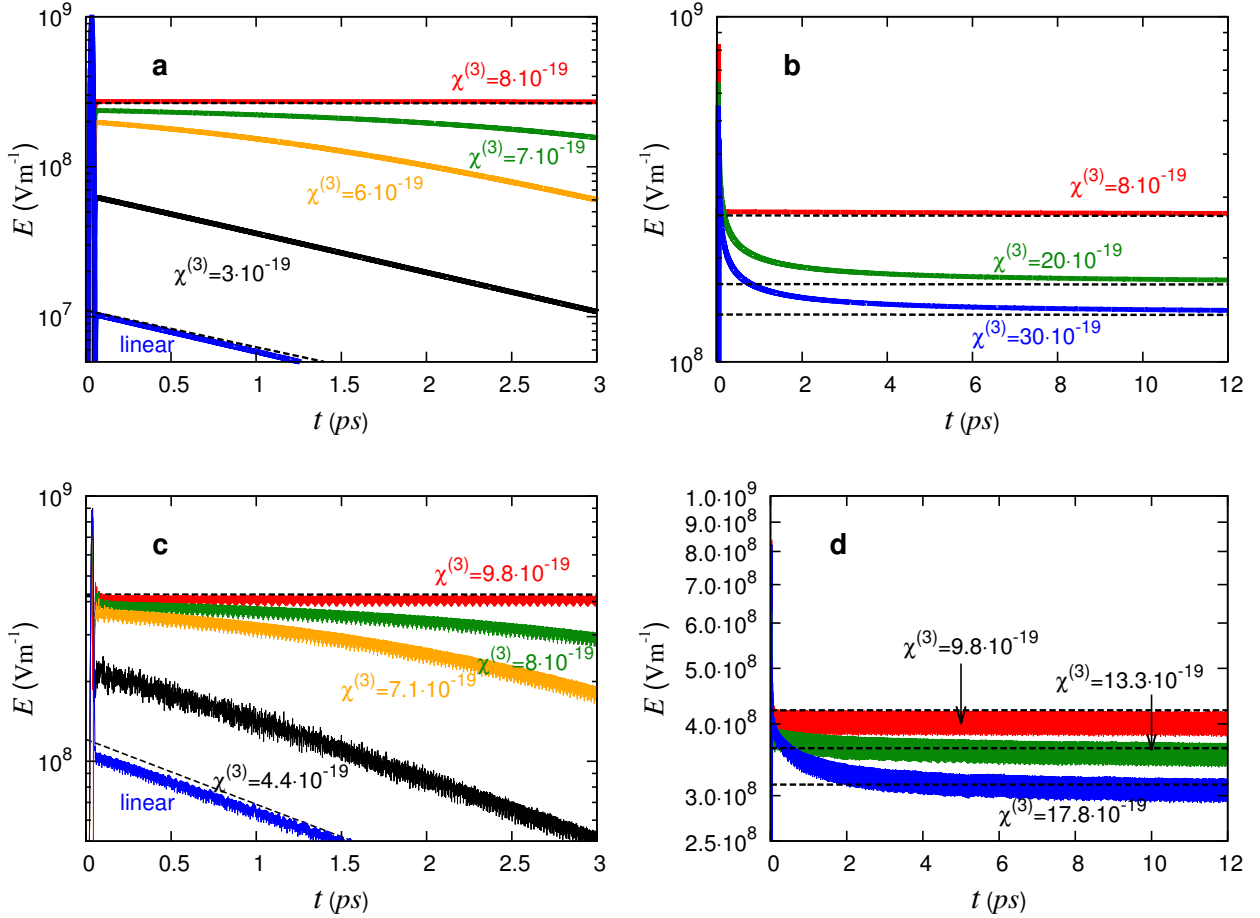, width=1\linewidth}
    \caption{\textbf{Excitation of the meta-atom in the nonlinear regime.} Influence of the value of $\chi^{(3)}$ (in $\meter^2\volt^{-2}$)
    on the trapped field decay for an incident pulse with $\Delta \tau=16.3 ~\femto\second$.
     (\textbf{a}) and (\textbf{b})  Analytical model. (\textbf{c}) and (\textbf{d}) CST Microwave Studio
     simulations.
     }
    \label{fig:CST_delta_nu_100}
\end{figure}

To illustrate the discussion, we show in Fig.
\ref{fig:CST_delta_nu_100} the dynamics of the peak electric field
obtained by numerically solving \eqref{E:differential_equation_time}
for different values of $\chi^{(3)}$. The structural parameters of
the meta-atom are as in the previous examples ($R_1 \approx 0.98
R_{1,0}$ and $R_2=1.1 R_1$), and the incident field amplitude is
 $E_0=10^9 ~\volt\meter^{-1}$. The incident pulse is
peaked at $t_0=0.047~\pico\second$ (sharp peak in Fig.
\ref{fig:CST_delta_nu_100}) and its duration is determined by
$\Delta \tau=16.3 ~\femto\second$. As seen in Fig.
\ref{fig:CST_delta_nu_100}a, as $\chi^{(3)}$ is increased the
external coupling efficiency is improved, whereas at the same time
the decay rate approaches the minimum value $\omega ''_{\rm{min}}
\approx \omega_\text{c} /2 \approx 0$. For $\chi^{(3)} \geq 8 \cdot
10^{-19}~\meter^2\volt^{-2}$ (Fig. \ref{fig:CST_delta_nu_100}b), the
field decay rate is $\omega''_{\rm{min}}$ and a radiation ``bit''
may stay trapped within the meta-atom for an extremely long period
of time. Consistent with Eq.
\eqref{E:expected_value_Chi_times_Esquare} this regime is
characterized by a specific value of $\chi^{(3)}|E_{\omega_\text{p}}|^2
\approx 0.056$. It is important to highlight that (if different from
zero) the trapped field steady-state amplitude only depends on the
strength of the nonlinearity, and not on the excitation.

\begin{figure}[!th]
\centering \epsfig{file=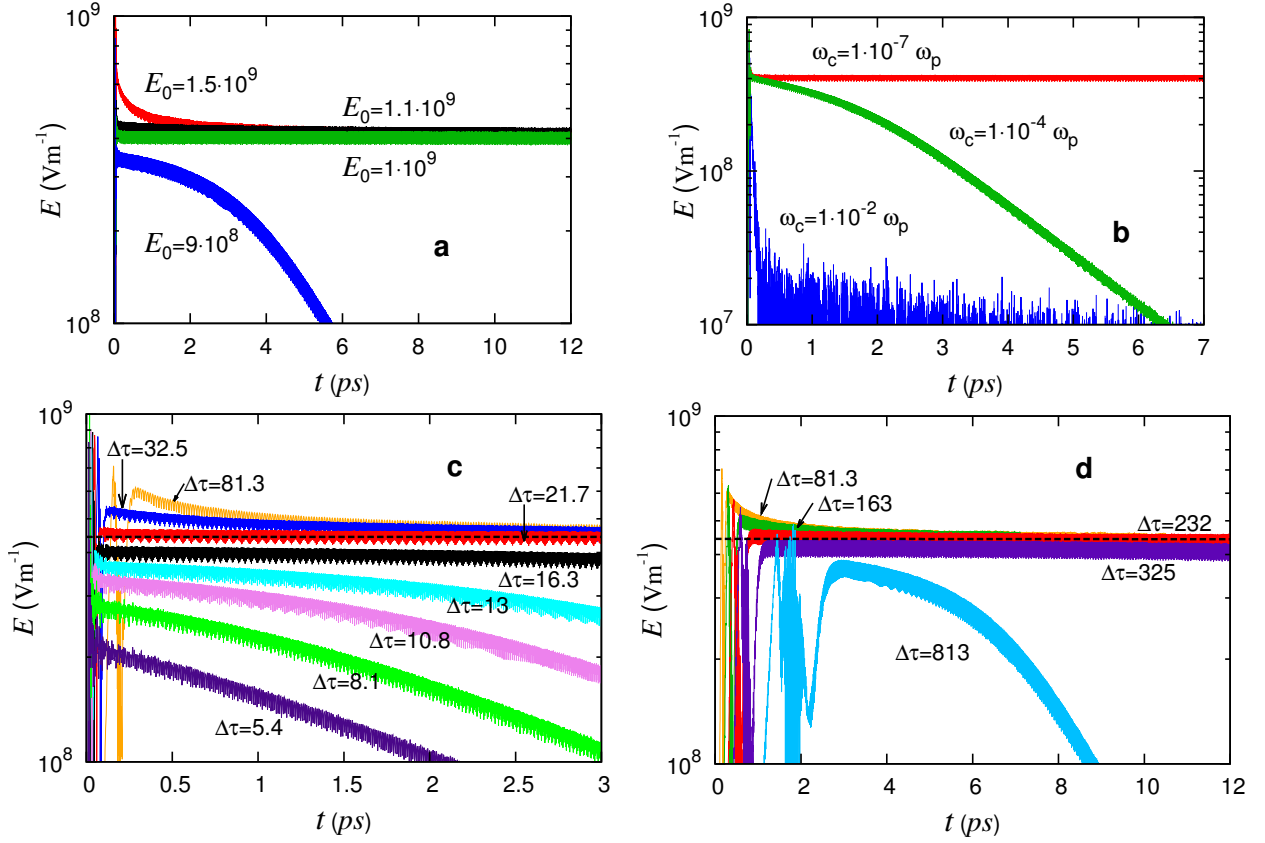, width=1\linewidth}
    \caption{\textbf{The effect of $E_0$, material loss, and pulse duration on the field decay.} Influence of (\textbf{a}) the incident field magnitude $E_0$ and (\textbf{b}) the level of losses $\omega_\text{c}$ on the trapped field decay
    for a core material with $\chi^{(3)}=9.8\cdot 10^{-19}~\meter^2\volt^{-2}$ and an incident pulse of duration $\Delta \tau=16.3 ~\femto\second$. Influence of the pulse duration (in \femto\second) on the trapped field decay
    for a core material with $\chi^{(3)}=8.89\cdot 10^{-19}~\meter^2\volt^{-2}$. (\textbf{c}) Short-pulse duration. (\textbf{d}) Long-pulse duration.}
    \label{fig:CST_chi3_1e-6}
\end{figure}

To validate our analytical model, we did a similar study using CST
Microwave Studio (Figs. \ref{fig:CST_delta_nu_100}c and d). As
seen, the full wave CST simulations are qualitatively consistent
with the results of the analytical model. Now, the radiation loss is
suppressed for $\chi^{(3)} \geq 9.8 \cdot
10^{-19}~\meter^2\volt^{-2}$. The threshold for light trapping is
$\chi^{(3)}|E_{\omega_\text{p}}|^2 \approx 0.175$, which is about three
times larger than the value predicted by the analytical model. The
threshold value can be made arbitrarily small by reducing the
detuning of $R_1$ with respect to $R_{1,0}$. Notably, for
$\chi^{(3)}=9.8 \cdot 10^{-19}~\meter^2\volt^{-2}$, the field
amplitude at the center of the meta-atom reaches, almost immediately
after the end of the incident pulse, the steady-state value
(indicated by the dashed horizontal lines in Fig.
\ref{fig:CST_delta_nu_100}b) determined by the quantized value of
$\chi^{(3)}|E_{\omega_\text{p}}|^2$. The electric field time animation for
the example of Fig. \ref{fig:CST_delta_nu_100}c with $\chi^{(3)}=9.8
\cdot 10^{-19}~\meter^2\volt^{-2}$ can be found in the supplementary
online materials (Supplementary Movie 1).  For a third-order
susceptibility greater than this value, the steady-state is only
reached after the extra-amount of electromagnetic energy is
released from the meta-atom. These results are confirmed by the
study in Fig. \ref{fig:CST_chi3_1e-6}a of the influence of the
incident pulse amplitude for a fixed value of the third-order
nonlinearity $\chi^{(3)}=9.8 \cdot 10^{-19}~\meter^2\volt^{-2}$. As
seen, for an amplitude above the threshold ($E_0=
10^9~\volt\meter^{-1} $) and after relaxation of the extra energy,
the field inside the core-shell particle always saturates at the
same level.

As expected, the presence of material loss deteriorates the
oscillation lifetime, but the results remain exciting for small
values of the collision frequency. This is illustrated by Fig.
\ref{fig:CST_chi3_1e-6}b which represents the field at the center of
the meta-atom for different levels of loss in the plasmonic shell.
Unfortunately, for a realistic level of metal loss it seems
unfeasible to hold the light within the meta-atom for a large number
of oscillation cycles. A solution that may help to alleviate this
problem is to mimic the $\varepsilon \approx 0$ regime using
dielectric photonic crystals operating in the vicinity of a band gap
edge. In general, a practical realization of our idea may require
some loss compensation mechanism
\cite{xiao_loss-free_2010,de_leon_amplification_2010}. Even though
challenging, this may be within reach with the current state of the
art technologies using either optical or electrical pumping
\cite{bergman_surface_2003,noginov_demonstration_2009,hill_lasing_2007,oulton_plasmon_2009,li_electric_2013,hill_advances_2014}.
Some encouraging results in this direction have been reported in the
recent literature, specifically the realization of loss-compensated
nanostructures relying on nanoscale gain media formed either by dye
molecules (e.g. rhodamine dye) or by semiconductor nanostructures
(e.g. quantum dots) such that the population inversion is created
optically or electrically
\cite{xiao_loss-free_2010,de_leon_amplification_2010,meinzer_arrays_2010}.

The external coupling efficiency depends on the pulse duration
$\Delta \tau$. To illustrate this, we represent in Fig.
\ref{fig:CST_chi3_1e-6}c and \ref{fig:CST_chi3_1e-6}d the inner
field dynamics determined with CST Microwave Studio, for the case
wherein the third-order susceptibility is kept constant $\chi^{(3)}=
8.89\cdot 10^{-19}~\meter^2\volt^{-2}$ and the \textit{fwhm} of the
incident pulse $\Delta \tau$  varies from $5.4$ to $813
~\femto\second$. As seen, for a pulse duration such that $5.4<\Delta
\tau<21.7~\femto\second$ the excitation efficiency is improved as
$\Delta \tau$ is increased, so that the field decay rate
progressively reaches the optimal value $\omega''_{\rm{min}} \approx
0$. For $21.7\leq\Delta \tau \leq 232~\femto\second$, the light can
be trapped inside the particle with suppressed radiation loss.
Again, for a pulse duration different from the threshold values
($21.7$ or $232~\femto\second$) the field does not reach immediately
the quantized value of $\chi^{(3)}|E_{\omega_\text{p}}|^2$ and some
extra-time is needed to release the excess of electromagnetic
energy. For $\Delta \tau>232~\femto\second$, the external coupling
efficiency is deteriorated and the temporal dynamics of the incident
pulse is incompatible with the requirements for light trapping.

\subsection*{Freeing the trapped light}

In the limit of no material loss the oscillation lifetime may be
extremely large (see Fig. \ref{fig:CST_chi3_1e-6}b), possibly
infinite. Parenthetically, we note that frequency conversion, e.g.
third harmonic generation, may also limit the oscillations lifetime.
Thus, from a theoretical point of view, it is interesting to discuss
how a trapped radiation ``bit'' may be released from the meta-atom.
We propose to do this using another Gaussian-shaped pulse. Hence,
next it is supposed that the meta-atom is sequentially illuminated
by two pulses. The first pulse has parameters
$t_{0,1}=0.047~\pico\second$, duration $\Delta
\tau_1=21.7~\femto\second$, and amplitude $E_{0,1}=10^9
~\volt\meter^{-1}$, and serves to trap a radiation ``bit'' in the
meta-atom. The core region is characterized by the susceptibility
$\chi^{(3)}=8.89\cdot 10^{-19}~\meter^2\volt^{-2}$.

One option to release the trapped radiation ``bit'' is to create a
collision between a second light pulse with the same oscillation
frequency $\omega_\text{p}$ and the meta-atom. We numerically verified that
this is indeed a valid strategy (not shown), but in practice it may
require very large field amplitudes (e.g. $E_{0,2} \sim 5 E_{0,1}$)
and the impact of the second pulse with the meta-atom leads to a
chaotic field behavior. To circumvent this problem, we imagined a
different solution relying on a light-pulse with the same amplitude
as the first one ($E_{0,2}=E_{0,1}$), but with a different
oscillation frequency $\omega_2$. The remaining parameters of the
second pulse are $t_{0,2} \approx 0.149~\pico\second$ and $\Delta
\tau_2 = 28.3~\femto\second$.
\begin{figure}[!ht]
\centering \epsfig{file=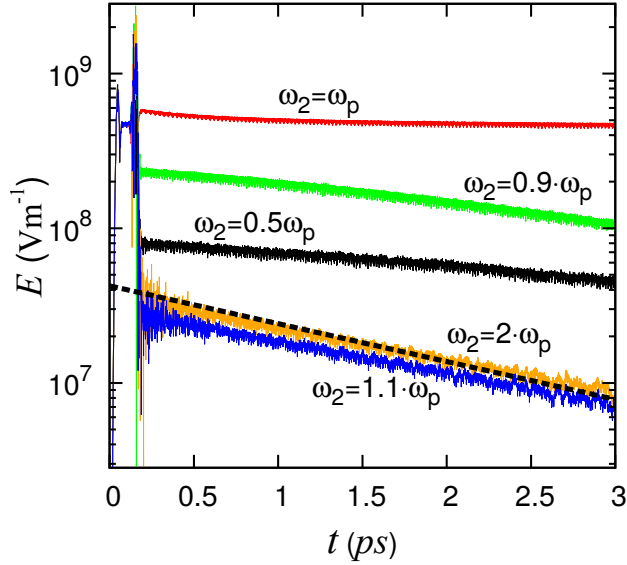, width=0.5\linewidth}
         \caption{\textbf{Release of the trapped light with a second light pulse.} Electric field as a function of
         time when the meta-atom is sequentially illuminated by two-light
         pulses. The collision of the second pulse with the
         meta-atom (sharp peak in the plots) releases the trapped radiation ``bit'', except
         when the frequency of oscillation of the second pulse satisfies $\omega_2 = \omega_\text{p}$. The dotted line
         represents the decay rate for the equivalent linear case.}
\label{fig:results_2_pulses}
\end{figure}
As seen in Fig. \ref{fig:results_2_pulses}, the first light-pulse
pumps the meta-atom ``embedded eigenstate''. Notably, when the
frequency of the second pulse $\omega_2$ is significantly different
from $\omega_\text{p}$, its collision with the meta-atom frees the
trapped radiation ``bit''. This occurs because mixing the two pulses
leads to both frequency and modal conversions. A time animation of
the collision for the example of Fig. \ref{fig:results_2_pulses}
with $\omega_2=1.1 \omega_\text{p}$ can be found  in the
supplementary materials (Supplementary Movie 2). Interestingly, when
$\omega_2>\omega_\text{p}$ the light release occurs with a decay
rate similar to the decay rate of the equivalent linear particle
(with $\chi^{(3)}=0$). Moreover, the field relaxation after the
second pulse overtakes the particle is faster for
$\omega_2>\omega_\text{p}$. Note also that for $\omega_2 =
\omega_\text{p}$ the radiation ``bit'' remains trapped within the
core after the collision with the second pulse.

\section*{Discussion}

In summary, we demonstrated that the interplay between a nonlinear
response and plasmonic effects may enable storing a quantized amount
of electromagnetic radiation in an open plasmonic resonator. We
investigated and characterized with full-wave simulations the
conditions required for the meta-atom self-tuning. It was found that
when the field amplitude and the pulse duration surpass certain
thresholds, a well defined amount of energy stays trapped in the
resonator. This regime with suppressed radiation loss is
characterized by a specific value of $\chi^{(3)}|E_{\omega_\text{p}}|^2$
close to the value predicted by Eq.
\eqref{E:expected_value_Chi_times_Esquare}. Hence, in a steady-state
the trapped energy is independent of the excitation.

In practice, the lifetime of the trapped radiation bit is always
limited by unavoidable material loss. One may envision that by
incorporating some form of electrical or optical gain
\cite{bergman_surface_2003,noginov_demonstration_2009,hill_lasing_2007,oulton_plasmon_2009,li_electric_2013,hill_advances_2014}
into the meta-atom it may be possible to compensate for the effects
of absorption and hold the trapped radiation within the
nano-resonator for a long period of time. The implementation of this
gain compensation mechanism is at the present time the main
challenge to render the light confinement in the optical meta-atom
experimentally observable. In this regard, one should note that even
if some gain element is incorporated into our proposal it remains
fundamentally different from a nanolaser. Indeed, provided the level
of gain in the linear regime is kept below the threshold associated
with lasing (i.e. the threshold required to over-compensate both the
material absorption and the radiation loss) the system remains
stable. The only way to trigger oscillations in the core region is
by using an external optical excitation. Crucially, due to the
nonlinear dynamics, the radiation loss of the system decreases when
sufficient energy is pumped into the resonator. Hence, the gain
required to compensate the total loss in the nonlinear regime is
expected to be less than the threshold gain that drives the system
into lasing in the linear regime. Because of this property, provided
the gain response does not saturate in the dynamic range of
interest, it may be feasible to guarantee a robust stable operation
of the meta-atom in the linear regime and a full compensation of
loss in the nonlinear regime. The combination of the meta-atom with
a gain element may thus provide a rudimentary one-bit optical
memory. It is relevant to mention that because the dimensions of the
meta-atom are of the order of the wavelength its resonances are well
separated in frequency. Thus, the presence of a gain medium with a
spectral gain response peaked at the frequency $\omega_\text{p}$ is
not expected to change in any manner the role of the secondary
resonances, which have much lower quality factors.

Very importantly, our ideas may be generalized to other type of
open-resonators, and unveil a novel mechanism to break the Lorentz
reciprocity principle and to efficiently couple high-Q optical
resonators with an external excitation. In particular, our solution
can be readily extended to more general geometries, not necessarily
based on core-shell particles. Indeed, a generic high-Q optical
resonator may be properly tuned such that when covered with an ideal
$\varepsilon=0$ shell it supports an embedded eigenvalue light state
with suppressed radiation loss. Notably, these structures are
expected to be much less sensitive to the effect of loss in the ENZ
cover, because for a non-uniform core the task of confining the
radiation within the meta-atom does not rely exclusively on the
$\varepsilon=0$ cover. Thus, such modified meta-atoms may determine
an exciting path towards the experimental verification of the light
trapping with quantized energy.

Remarkably, a collision between another light pulse with different
oscillation frequency and the meta-atom may enable releasing the
trapped radiation ``bit''. We envision that the developed ideas may
find applications in chemical or biological sensing, light emitting
sources, and optical memories.

\section*{Methods}

The numerical results shown in Figs. \ref{fig:linear_results} to
\ref{fig:results_2_pulses} were obtained using the time domain
solver of the commercial software CST Microwave Studio, using open
boundary conditions and plane-wave excitation. We numerically
checked that our $\chi^{\left( 3 \right)}$ parameter is related to
the definition adopted by CST as: $\chi ^{\left( 3 \right)} =
\frac{8}{9} \chi^{\left( 3 \right)}_\text{CST}$.

%%%%%%%%%%%%%%%%%%%%%%%%%%%%%%%%%%%%%%%%%%%%%%%%%%%%%%%%%%%%%%%%%%%%%%%%%%%%%%%%%%%%%%%%%%%%%%%%%%%%%%%
%%%%%%%%%%%%%%%%%%%%%%%%%%%%%%%%%%%%%%%%%%%%%%%%%%%%%%%%%%%%%%%%%%%%%%%%%%%%%%%%%%%%%%%%%%%%%%%%%%%%%%%

% \bibliographystyle{naturemag}
%
% \bibliography{Biblio_vfinal}

\begin{acknowledgments}
This work was partially funded by Funda\c{c}\~{a}o para a
Ci\^{e}ncia e a Tecnologia under projects PTDC/EEITEL/2764/2012,
PTDC/EEITEL/4543/2014, and UID/EEA/50008/2013 and by Instituto de
 Telecomunica\c{c}\~{o}es under project C00355-TRAP.
\end{acknowledgments}

\section*{Author contributions}
S.L. carried out the analytical modelling and numerical simulations.
M.S. conceived and supervised the work. Both authors discussed the
theoretical and numerical aspects and interpreted the results. Both
authors contributed to the preparation and
writing of the manuscript.\\

\noindent \textbf{Competing financial interests:} The authors declare no competing financial interests.\\

\section*{Figure Legends}

Fig.1: \textbf{The optical meta-atom.} The core material is a dielectric with a
         nonlinear permittivity response $\eps_1^{\rm{NL}}$, while the shell is a plasmonic material with permittivity $\eps_2(\omega)$.
         The inner  and outer radii are $R_1$ and $R_2$, respectively.\\

Fig.2: \textbf{Excitation of the meta-atom in the linear regime.} Electric field
         at the center of the meta-atom as a function of time. The
         incident pulse duration is $\Delta \tau=16.3 ~\femto\second$.
         (\textbf{a}) $R_1 \approx 0.98 R_{1,0}$.
         The black thin line represents the theoretical peak amplitude determined by the decay rate $\omega''_\text{r}$.
         (\textbf{b})  $R_1=R_{1,0}$.\\

Fig.3: \textbf{Excitation of the meta-atom in the nonlinear regime.} Influence of the value of $\chi^{(3)}$ (in $\meter^2\volt^{-2}$)
    on the trapped field decay for an incident pulse with $\Delta \tau=16.3 ~\femto\second$.
     (\textbf{a}) and (\textbf{b})  Analytical model. (\textbf{c}) and (\textbf{d}) CST Microwave Studio
     simulations.\\

Fig.4: \textbf{The effect of $E_0$, material loss, and pulse duration on the field decay.} Influence of (\textbf{a}) the incident field magnitude $E_0$ and (\textbf{b}) the level of losses $\omega_\text{c}$ on the trapped field decay
    for a core material with $\chi^{(3)}=9.8\cdot 10^{-19}~\meter^2\volt^{-2}$ and an incident pulse of duration $\Delta \tau=16.3 ~\femto\second$. Influence of the pulse duration (in \femto\second) on the trapped field decay
    for a core material with $\chi^{(3)}=8.89\cdot 10^{-19}~\meter^2\volt^{-2}$. (\textbf{c}) Short-pulse duration. (\textbf{d}) Long-pulse duration.\\

Fig.5: \textbf{Release of the trapped light with a second light pulse.} Electric field as a function of
         time when the meta-atom is sequentially illuminated by two-light
         pulses. The collision of the second pulse with the
         meta-atom (sharp peak in the plots) releases the trapped radiation ``bit'', except
         when the frequency of oscillation of the second pulse satisfies $\omega_2 = \omega_\text{p}$. The dotted line
         represents the decay rate for the equivalent linear case.\\

\end{document}